\documentclass{aa}

\usepackage{graphicx}
\usepackage{txfonts}
\usepackage[normalem]{ulem}
\usepackage[colorlinks=true, linkcolor=blue, citecolor=blue]{hyperref}
\AtBeginShipout{%
	\ifnum\value{page}>1 %
	\typeout{* Additional boxing of page `\thepage'}%
	\setbox\AtBeginShipoutBox=\hbox{\copy\AtBeginShipoutBox}%
	\fi
}
\begin{document} 

\title{Optimizing the subwavelength grating of L-band annular groove phase masks for high coronagraphic performance}

\titlerunning{High-performance annular groove phase masks}


\author{E.~Vargas Catal\'an \inst{1}
\and E.~Huby \inst{2}
\and P.~Forsberg \inst{1}
\and A.~Jolivet \inst{2}
\and P.~Baudoz \inst{3}
\and B.~Carlomagno \inst{2}
\and C.~Delacroix \inst{4}
\and S.~Habraken \inst{2}
\and D.~Mawet \inst{5,6}
\and J.~Surdej \inst{2}
\and O.~Absil \inst{2}\fnmsep\thanks{F.R.S.-FNRS Research Associate}
\and M.~Karlsson \inst{1}
}

\institute{Department of Engineering Sciences, {\AA}ngstr\"{o}m Laboratory, Uppsala University, Box 534, 751 21 Uppsala, Sweden
\\
\email{mikael.karlsson@angstrom.uu.se}
\and Space sciences, Technologies and Astrophysics Research (STAR) Institute, Universit\'e de Li\`ege, 19c All\'ee du Six Ao\^ut, B-4000 Li\`ege, Belgium
\and LESIA-Observatoire de Paris, CNRS, UPMC Univ.\ Paris 06, Univ.\ Paris-Diderot, 5 pl.~J.~Janssen, F-92195 Meudon, France
\and Sibley School of Mechanical and Aerospace Engineering, Cornell University, Ithaca, NY 14853, USA
\and Department of Astronomy, California Institute of Technology, 1200 E. California Blvd, MC 249-17, Pasadena, CA 91125, USA
\and Jet Propulsion Laboratory, California Institute of Technology, 4800 Oak Grove Drive, Pasadena, CA 91109, USA
}

\date{Received 10 April 2016 / Accepted 4 August 2016}

 
\abstract
{The annular groove phase mask (AGPM) is one possible implementation of the vector vortex coronagraph, where the helical phase ramp is produced by a concentric subwavelength grating. For several years, we have been manufacturing AGPMs by etching gratings into synthetic diamond substrates using inductively coupled plasma etching.}
{We aim to design, fabricate, optimize, and evaluate new L-band AGPMs that reach the highest possible coronagraphic performance, for applications in current and forthcoming infrared high-contrast imagers.}
{Rigorous coupled wave analysis (RCWA) is used for designing the subwavelength grating of the phase mask. Coronagraphic performance evaluation is performed on a dedicated optical test bench. The experimental results of the performance evaluation are then used to accurately determine the actual profile of the fabricated gratings, based on RCWA modeling.}
{The AGPM coronagraphic performance is very sensitive to small errors in etch depth and grating profile. Most of the fabricated components therefore show moderate performance in terms of starlight rejection (a few 100:1 in the best cases). Here we present new processes for re-etching the fabricated components in order to optimize the parameters of the grating and hence significantly increase their coronagraphic performance. Starlight rejection up to 1000:1 is demonstrated in a broadband L filter on the coronagraphic test bench, which corresponds to a raw contrast of about $10^{-5}$ at two resolution elements from the star for a perfect input wave front on a circular, unobstructed aperture.}
{Thanks to their exquisite performance, our latest L-band AGPMs are good candidates for installation in state of the art and future high-contrast thermal infrared imagers, such as METIS for the E-ELT.}

\keywords{Instrumentation: high angular resolution -- Planetary systems -- Planets and satellites: detection}

\maketitle
\section{Introduction}

So far, most exoplanets have been detected by indirect methods, based on the measurement of the effect of the companion on its host star, either in its spectrum thanks to the Doppler effect or in its photometric curve during transits. The currently known exoplanet population is therefore biased, since these techniques are sensitive to relatively short period and very close massive companions. In that context, direct imaging offers a good complement to probe larger separations around stars. In addition, the direct detection of photons emitted or reflected by the planet allows for photometric and spectroscopic studies, which is crucial to get insight into its atmospheric composition. It also enables precise astrometry, which over time provide orbital characteristics of the planets and insight into the dynamical environment and history. However, direct imaging is a challenging technique, as it requires to reach a very high contrast, typically ranging from $10^{-4}$ to $10^{-10}$ for hot giant planets to Earth-like planets, respectively, and a high angular resolution ($\sim0\farcs1$). Coronagraphy, which aims to reject the glaring light of the central star to enhance the signal from the faint companion, combined with extreme adaptive optics systems and advanced image processing, is a requirement to reach such performance.

Among all possible coronagraph designs, the vortex coronagraph was proposed a decade ago by \cite{Mawet05a} and \cite{Foo05}. It consists of a focal plane phase mask inducing a phase ramp around the optical axis. When passing through the phase mask, the light of an on-axis star is diffracted and redistributed outside the geometrical pupil of the telescope in a downstream pupil plane. A diaphragm, referred to as Lyot stop, is then used to block the light of the central star. The light of an off-axis companion is not, or only partially, affected by the vortex phase pattern and can propagate towards the detector.

One possible implementation of vortex phase masks is based on the manufacturing of concentric rings creating a subwavelength rotational grating, in a design referred to as the annular groove phase mask \citep[AGPM,][]{Mawet05a}. For a given linear polarization of the incoming light, the phase mask acts as a rotating phase retarder, thus inducing the desired phase ramp. Our team has previously demonstrated a fabrication process for diamond AGPMs working in the L (3.4--4.1~$\mu$m), M (4.4--5.0~$\mu$m) and N (10--13~$\mu$m) bands \citep{Forsberg13}. The diamond AGPMs were manufactured using nanoimprint lithography (NIL) and inductively coupled plasma reactive ion etching (ICP-RIE) in high density plasmas using highly oxidizing chemistries \citep{Karlsson03, Hwang04, Gu2004}. The fabricated AGPMs designed for the L band were characterized on the YACADIRE coronagraphic test bench at Observatoire de Paris \citep{Delacroix13}, showing starlight rejection up to 500:1. Our phase masks now equip infrared coronagraphs on several 10-m class telescopes (VLT/NACO, \citealt{Mawet13}; VLT/VISIR, \citealt{Del12b}; LBT/LMIRCam, \citealt{Defrere14}; Keck/NIRC2, Serabyn et al., submitted to \aj).

Here we report on a new generation of AGPMs optimized for the highest possible coronagraphic performance in the L and M bands. The design of the subwavelength grating based on rigorous coupled wave analysis \citep[RCWA,][]{Mawet05b} is described in Sect.~\ref{sec:design}. A short description of our improved fabrication process for high aspect ratio diamond gratings, and of means to assess the grating parameters during and after etching is then given in Sect.~\ref{sec:fabrication}. Section~\ref{sec:performance} focuses on the performance assessment of the phase masks using the YACADIRE coronagraphic test bench at the Observatoire de Paris. Considering that the depth of the grating is a determining parameter, we propose in Sect.~\ref{sec:tuning} two possible methods to finely tune the grating depth with further etching and thereby reach the best possible coronagraphic performance. In Sect.~\ref{sec:newperf}, this process is demonstrated on a few AGPMs, which have been successfully re-etched and show significantly improved coronagraphic performance after tuning.

\begin{figure}
\centering
\includegraphics[width=3.6cm]{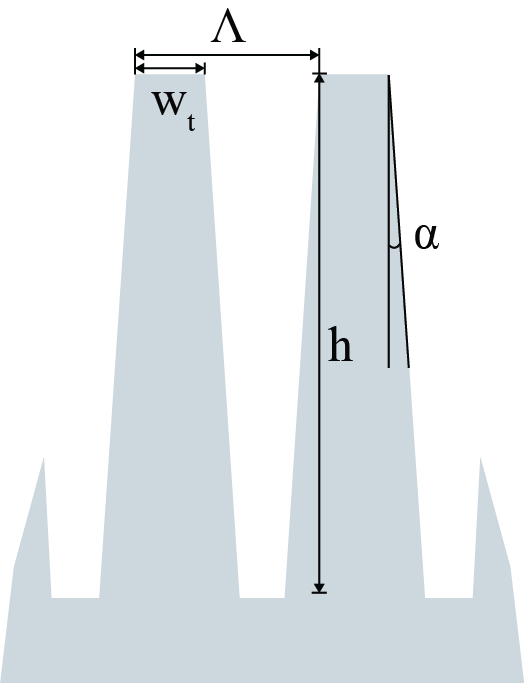}
\caption{Schematic picture of a cross section of the AGPM, showing the sidewall angle ($\alpha$), the grating depth ($h$) the line width ($w_t$) and the grating period ($\Lambda$).}
\label{fig:agpm}
\end{figure}

\begin{figure}
\centering
\includegraphics[width=7.5cm]{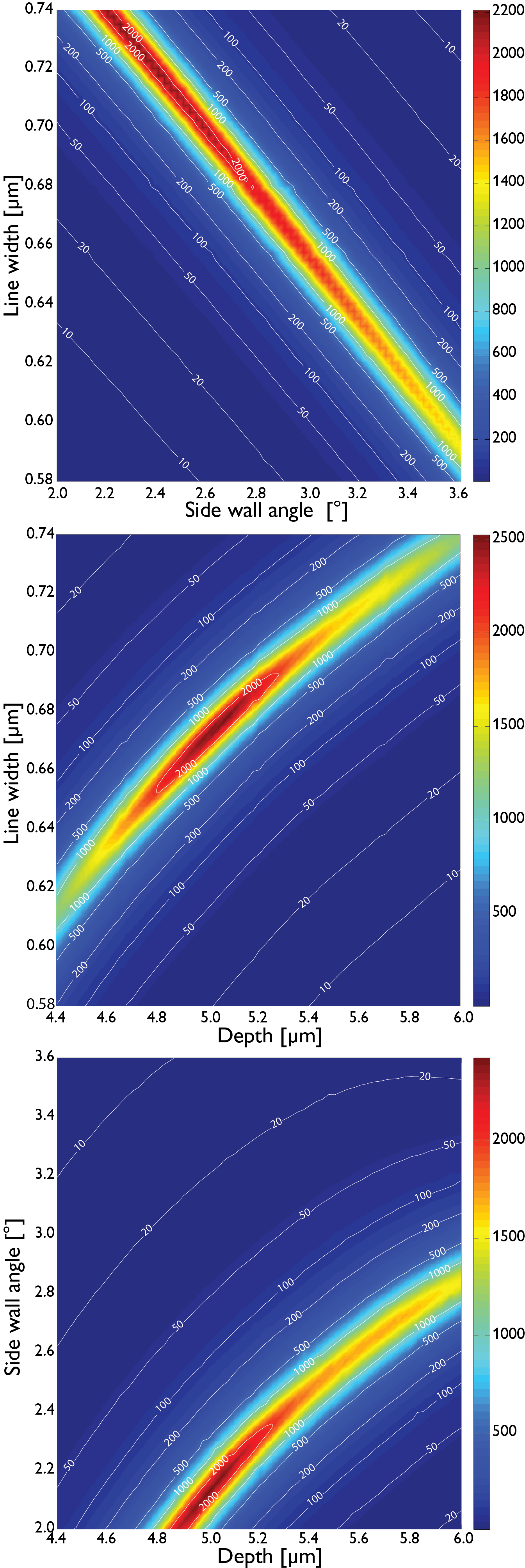}
\caption{RCWA simulations showing the rejection ratio for a grating period of 1.42~$\mu$m on a 3.4-4.1~$\mu$m broadband filter. \textit{Top.} Fixed grating depth of 5.5~$\mu$m as a function of the line width $w_t$ and sidewall angle $\alpha$. \textit{Middle.} Fixed sidewall angle of $2\fdg45$ as a function of the grating depth $h$ and line width $w_t$. \textit{Bottom.} Fixed line width of 0.7~$\mu$m, as a function of the sidewall angle $\alpha$ and grating depth $h$.}
\label{fig:rcwa}
\end{figure}

\section{Design and simulation} \label{sec:design}

\begin{figure*}
\centering
\includegraphics[width=18cm]{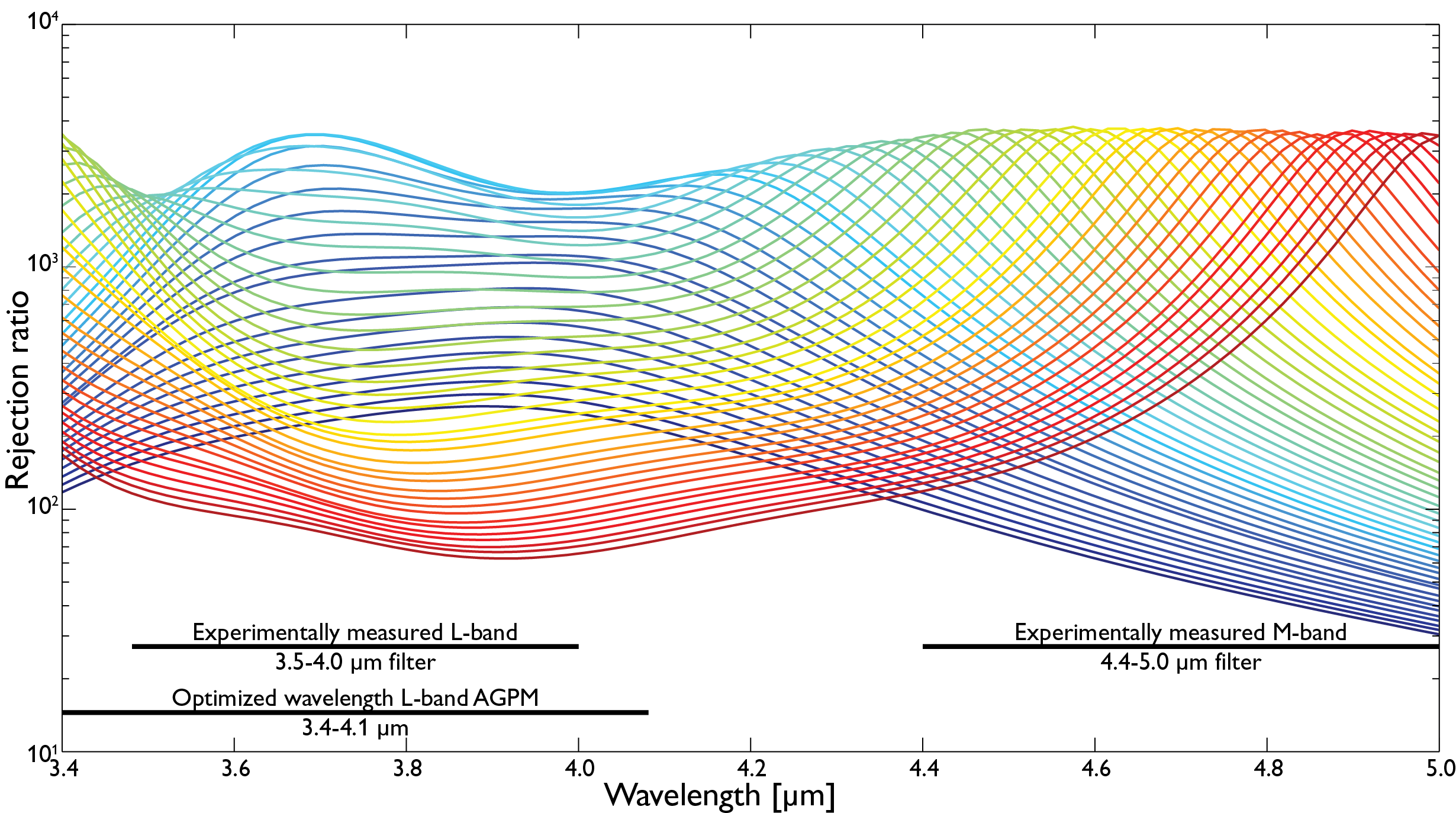}
\caption{Rejection ratio as a function of wavelength for different grating depths between 5.0 and 6.0~$\mu$m (from blue to red), with the lines separated by steps of 0.02~$\mu$m. The sidewall angle is set to $2\fdg45$ and the line width $w_t$ is 0.70~$\mu$m.}
\label{fig:rejection_profile}%
\end{figure*}

The subwavelength grating composing the AGPM \citep{Mawet05b} is defined through the following parameters (see Fig.~\ref{fig:agpm}):  the grating period $\Lambda$, the line width at the top of the grating $w_t$, the depth of the grating $h$, and the angle of the sidewall $\alpha$. The grating period $\Lambda$ is fixed to fulfill the subwavelength criterion: $\Lambda < \lambda/n$ (when the ambient medium is air), where $\lambda$ is the illuminating wavelength and $n$ is the refractive index of the substrate ($n = 2.38$ for diamond in the thermal infrared regime). Inserting the values means that $\Lambda < 1.428$~$\mu$m for the short-wave end of the L band ($\lambda=3.4$~$\mu$m). Here, we set $\Lambda$ to 1.42~$\mu$m for all our L-band components. All our AGPMs feature a two-dimensional subwavelength grating on their back side, acting as an anti-reflection treatment. This grating keeps internal reflections at the interface between diamond and air below 2\%, which effectively reduces double-pass ghost signals in our AGPMs to less than 0.1\% over the whole L band \citep{Delacroix13}.

The sidewall angle $\alpha$ is determined by the etch process and needs to be low in order to reach high aspect ratios (grating depth divided by the width of the grooves). The process used to make most AGPMs presented in this paper produces a sidewall angle close to $2\fdg45$. With $\Lambda$ and $\alpha$ known, we used RCWA simulations to find the optimal values of $w_t$ and $h$ for high coronagraphic performance \citep{Delacroix12}. The starlight rejection efficiency is quantified by the \emph{rejection ratio} $R$, defined as the ratio of the total intensity of the non-attenuated point spread function (PSF) to the total intensity of the PSF attenuated by the coronagraph. In Fig.~\ref{fig:rcwa}, the rejection ratio is plotted with respect to a known period and one constant parameter as a function of the other two. From this figure, it becomes evident that, to a large extent, an error in the line width can be compensated by changing the etch depth (and vice versa) to improve the rejection ratio. The simulations also show that a variation of $0\fdg1$ in $\alpha$ can lead to a significant loss of performance. Moreover, the RCWA simulations presented in Fig.~\ref{fig:rcwa} reveal that a very small change in $w_t$ (by $\sim10$~nm) or $h$ (by $\sim100$~nm) can lead to a dramatically lowered optical performance of the AGPM. Figure~\ref{fig:rejection_profile} illustrates the fundamental limitations to the rejection ratio of AGPM-based coronagraphs on broadband filters. While the mean rejection ratio on a broadband L filter (3.4--4.1~$\mu$m) could reach up to 2500:1, an AGPM covering both L and M bands (3.4--5.0~$\mu$m) cannot reach a rejection ratio larger than 200:1 simultaneously in both bands. Because of the uncertainties in the etching process, we consider a rejection ratio of about 500:1 to be the maximum value we can reach on a broadband L filter using a single etching process, based on our previous fabrication attempts \citep{Delacroix13}.

\section{Fabrication and grating characterization} \label{sec:fabrication}

\begin{table*}
\caption{Grating parameters, expected and measured broadband rejection ratios at L band (3.5--4.0 $\mu$m) and M band (4.4--5.0 $\mu$m) for our AGPMs after fabrication.}
\label{tab:agpm_params}      
\centering
\begin{tabular}{l c c c c c c c} 
\hline\hline       
 & $w_t$ & $h$ & $\alpha$ & Expected $R$ & Measured $R$ & Expected $R$ & Measured $R$ \\ 
\multicolumn{1}{c}{Name} & [$\mu$m] & [$\mu$m] & [degrees] & (L band) & (L band) & (M band) & (M band) \\
\hline                    
AGPM-L5 & $0.630\pm 0.015$ & $4.90\pm 0.20$ & $3.20\pm 0.20$ & 30--2400 & 550 & 10--130 & 80 \\ 
AGPM-L6\tablefootmark{a} & $0.625\pm0.015$ & $4.57\pm0.20$ & $3.20\pm 0.20$ & 20--1600 & 150 & 10--50 & 30 \\
AGPM-L7\tablefootmark{a} & $0.625\pm0.015$ & $4.82\pm0.20$ & $3.45\pm 0.20$ & 20--1400 & 550 & 10--110 & N/A \\
AGPM-L8 & $0.645\pm0.015$ & $4.86\pm 0.20$ & $3.45\pm0.20$ & 30--270 & 50 & 10--80 & N/A \\
AGPM-L9\tablefootmark{b} & $0.750\pm0.010$ & $5.10\pm0.05$ & $2.45\pm0.10$ & 20--110 & 30 & 10--40 & 20 \\
AGPM-L10\tablefootmark{b} & $0.630\pm0.010$ & $5.20\pm0.05$ & $2.10\pm0.10$ & 10--40 & 20 & 70--450 & 90 \\
AGPM-L11\tablefootmark{b} & $0.650\pm0.010$ & $4.92\pm0.05$ & $2.22\pm 0.10$ & 40--670 & 70 & 110--520 & 240 \\
AGPM-L12\tablefootmark{b} & $0.650\pm0.010$ & $4.92\pm0.05$ & $2.22\pm 0.10$ & 40--670 & 70 & 110--520 & 120 \\
AGPM-L13\tablefootmark{b} & $0.590\pm0.010$ & $4.67\pm0.05$ & $2.45\pm0.10$ & 30--250 & 110 & 70--250 & 150 \\
AGPM-L14\tablefootmark{c} & $0.615\pm0.010$ & $4.67\pm0.05$ & $2.45\pm0.10$ & 70--1860 & 370 & 50--160 & 90 \\
AGPM-L15 & $0.630\pm0.010$ & $4.67\pm0.05$ & $2.45\pm0.10$ & 130--2300 & 630 & 40--120 & 70 \\
\hline                  
\end{tabular}
\tablefoot{
\tablefoottext{a}{Installed in the Keck/NIRC2 camera.}
\tablefoottext{b}{Chosen for grating tuning demonstration.}
\tablefoottext{c}{Installed in the LBT/LMIRCam camera.}}
\end{table*}

Polycrystalline diamond substrates of optical quality (Diamond Materials GmbH and Element Six Ltd.) with a diameter of 10~mm and a thickness of 300~$\mu$m were used. We have recently demonstrated an improved fabrication process for high aspect ratio diamond gratings \citep{Vargas16}. Most of the AGPMs presented in this work have been manufactured using this process, which involves nano-replication and ICP-RIE of Al, Si and diamond. Previously we used NIL in the nano-replication step \citep{Forsberg13,Delacroix13}, but we noticed that this process gave rise to a large reduction in line width, and that variations in line width were common, especially around the center of the AGPM. In our new process, we use solvent assisted micro molding (SAMIM, \citealt{Kim97, Vargas16}), which gives very good fidelity in the replicated patterns with nearly no difference in line widths compared to the master AGPM pattern. Moreover, our improved fabrication process use pure oxygen chemistry during the ICP-RIE of diamond, yielding a lower sidewall angle \citep{Vargas16}, which is beneficial for fabricating high performing AGPMs.

Eleven AGPMs were successfully fabricated (see Table~\ref{tab:agpm_params}). They were numbered from AGPM-L5 to L15 \citep[AGPM-L1 to L4 were presented in][]{Delacroix13}. We would like to point out that AGPM-L5 to L8 were fabricated by using our first generation of fabrication process based on NIL in the nano-replication step \citep{Forsberg13,Delacroix13}, except that C$_4$F$_8$ was added as an etch gas in the Si etch step giving less shrinkage in line width \citep{Vargas16}. When using the first generation of fabrication process, we had to repeat the nano-replication and thin film etching steps several times (for a given substrate) before getting an AGPM with correct line width (acceptable values: 590~nm~$\leq w_t \leq$~750 nm). As a result, the surface of the diamond substrate was degraded and for this reason, we had to discard several diamond substrates. Using our improved process completely removes these problems.

Evaluating the parameters of the etched gratings is not an easy task. Indeed, metrological methods such as atomic force microscopy (AFM) cannot reach down the trenches, and the features are too small for optical interferometers. Furthermore, a precise value of the sidewall angle can only be measured by cracking the AGPM perpendicularly to the grating to resolve a cross section of the profile in scanning electron microscopy (SEM). Therefore, the geometry of the AGPMs' high aspect ratio gratings was analyzed by cross section micrographs using SEM. However, since all of the AGPMs are potentially to be installed in telescopes, none was cracked except AGPM-L10. For each batch of diamond AGPMs (i.e., AGPMs etched together and therefore having almost the same sidewall angle), a twin sample was cracked instead. The twin is a test sample that follows the batch through the complete process. It was measured after each critical step. The grating parameters $w_t$ and $\alpha$ are indeed known to vary during the fabrication process; hence it is critical to follow the process using a twin sample, enabling recalculations of the optimal depth $h$.

The twin sample was cracked after the first Al etching step to check if the pattern was transferred successfully, and after the second Al etching to see if the mask layers have smooth sidewalls and to measure the line width before etching. The optimal etch depth was then recalculated by RCWA simulations, using the measured line width and sidewall angle. A third cracking was performed just before reaching the optimal etch depth (based on etch time, using the mean value of the diamond etch rate), to avoid too deep gratings. Previous etch runs showed that the diamond etch rate can vary up to 5\% \citep{Vargas16}. For the AGPMs demonstrated in \cite{Delacroix13} and AGPM-L5 to AGPM-L8, we wrongly assumed that the etch rate was always the same for our diamond etch recipe, thus giving a larger error in final etch depth (and in $w_t$ and $\alpha$) compared to using a twin sample. Again, $w_t$, $\alpha$ (and $h$) were measured and a final RCWA calculation was made for deciding on the optimal etch depth $h$. In the final step, the grating generally just needed to be etched 100--400~nm deeper to reach the optimal depth. The twin was cracked for a final time; $w_t$, $\alpha$ and $h$ were determined for the twin, and the parameter values for the AGPM fabricated in parallel were assumed to be the same. The measured grating parameters are reported in Table~\ref{tab:agpm_params}.


\section{Coronagraphic performance evaluation} \label{sec:performance}

The AGPMs were optically tested on the YACADIRE testbench at LESIA, Observatoire de Paris \citep{Boccaletti08}. This bench was previously used to characterize our first generation of AGPMs using a broadband L filter \citep{Delacroix13}. We refer to these two papers for a detailed description of the bench. On YACADIRE, the entrance pupil is defined by a circular aperture. The AGPM is placed at the focal plane, where the beam is converging at $f/40$, resulting in a diffraction pattern of full width at half maximum FWHM~$\simeq 150$~$\mu$m at L band. The diameter of the Lyot stop is undersized to 80\% of the original pupil size.

\begin{figure}
\centering
\includegraphics[width=8.8cm]{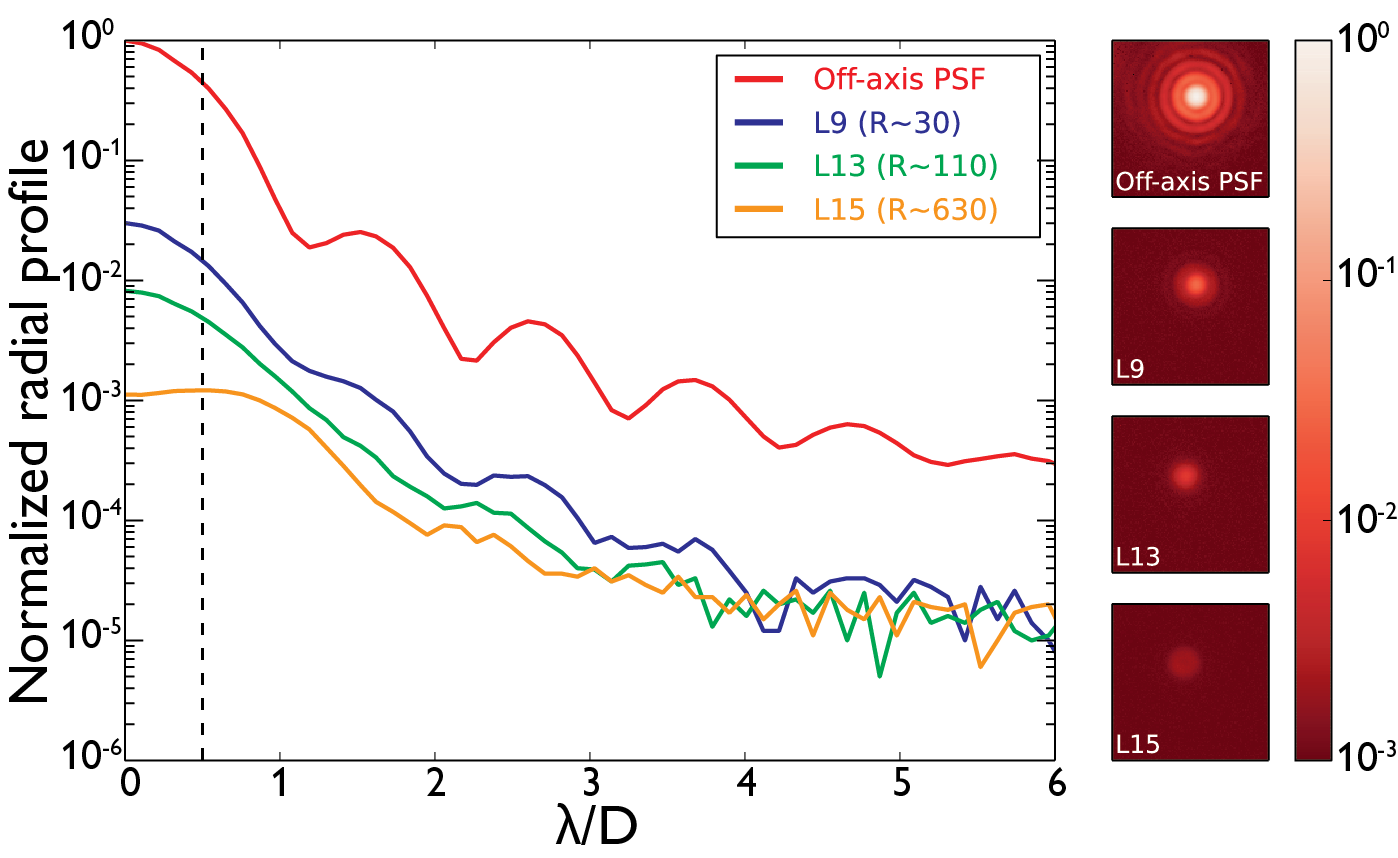}
\caption{Typical experimental results of AGPM optical performance characterization. \textit{Left}. Radial profile of the image with an AGPM translated by 1 mm (off-axis PSF), and with three different centered AGPMs, showing low (AGPM-L9), median (AGPM-L13) or high (AGPM-L15) performance after initial etching. The vertical dashed line shows the limit of the disk on which the flux is integrated to compute the rejection ratio $R$. \textit{Right}. Corresponding images shown with a logarithmic scale.}
\label{fig:agpm_psf}
\end{figure}

\begin{figure*}
\centering
\includegraphics[width=18cm]{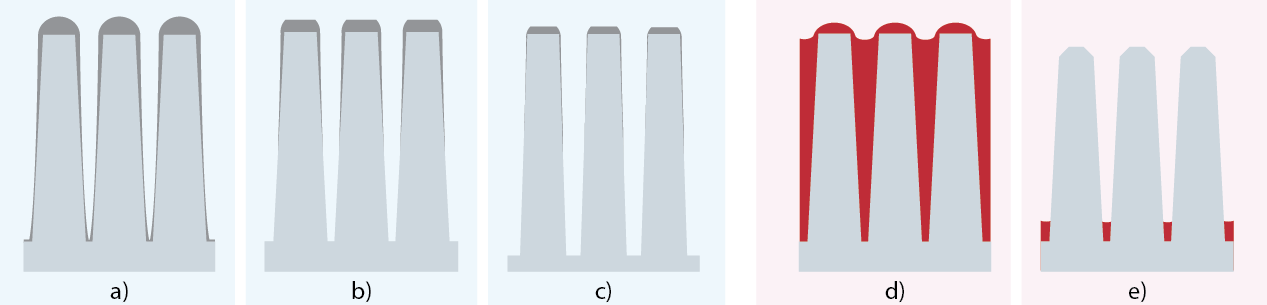}
\caption{Schematic of the tuning process. The three left-most sketches (with a blue background) show the process steps for deeper gratings; a) Al deposition, b) Al etching and c) diamond etching. The two right-most sketches (red background) show the process steps for shallower gratings; d) resist filling and e) diamond etching.}
\label{fig:tuning_process}
\end{figure*}

While the theoretical rejection ratio computed in RCWA simulations corresponds to the ratio of the total intensity in the two PSFs, this metrics is not practical in the case of our experimental data for two main reasons. First, the large thermal background encountered on the non-cryogenic YACADIRE bench reduce the exploitable part of the coronagraphic PSF to an angular separation of about $4\lambda/D$ (see Fig.~\ref{fig:agpm_psf}). Beyond this separation, the signal becomes dominated by background noise and by residuals associated to the background subtraction process. Second, the YACADIRE bench is not free from optical aberrations. Indeed, when the AGPM intrinsic rejection ratio exceeds 100:1, one can notice significant deformation of the coronagraphic intensity profile compared to the non-coronagraphic profile, and in particular a bump appearing at about $1\lambda/D$ (see Fig.~\ref{fig:agpm_psf}). This behavior is consistent with low-order aberrations dominating the coronagraphic performance. The vortex effect associated to the AGPM only affects the coherent part of the input beam, and reveals these low-order aberrations that were unnoticeable in non-coronagraphic images. In order to assess the true performance of the AGPM, we propose to compute the experimental rejection ratio $R$ by integrating the flux on an aperture of size equal to the resolution element $\lambda/D$, which encircles 80\% of the total energy in the non-coronagraphic PSF, and where the contribution of the coherent core is most prominent. This definition of the rejection ratio is the same as the one used in \citet{Delacroix13}, and would be strictly equivalent to the definition used in the RCWA simulations of Sect.~\ref{sec:design} if the optical system was perfect. Due to noise and optical aberrations, the measured rejection ratios will generally be somewhat underestimated, especially for the highest rejection ratios.

Our rejection ratio measurements were done using various spectral filters (see Table~\ref{tab:filters}): broadband L or M filters were placed directly in the cryostat to reduce background emission, while narrow-band filters were used at room temperature. The coronagraphic performance measurements performed in the two broadband filters are reported in Table~\ref{tab:agpm_params}. As expected, a large fraction of the fabricated AGPMs do not reach a rejection ratio of 100, which we consider as a bare minimum for on-sky coronagraphic applications \citep{Mawet10}.

\begin{table}
\caption{Central wavelength and width of the filters used for the AGPM coronagraphic performance evaluation.} 
\label{tab:filters}
\centering          
\begin{tabular}{c c c } 
\hline\hline       
Filter & $\lambda_0$ [$\mu$m] & $\Delta \lambda$ [$\mu$m] \\
\hline                    
Broad L & 3.750 & 0.50 \\ 
Narrow L-short & 3.475 & 0.10 \\
Narrow L-mid & 3.800 & 0.18 \\
Narrow L-long & 4.040 & 0.16 \\
Broad M & 4.700 & 0.60 \\
\hline                  
\end{tabular}
\end{table}


\section{Grating tuning processes} \label{sec:tuning}

Using the results from the performance measurements together with the RCWA calculations, and using the grating parameters measured on the twin sample as a first guess in the RCWA modeling, it becomes possible to determine more precisely the parameters of the grating, and to compute by how much the grating parameters need to be tuned to improve the AGPM coronagraphic performance. While the line width $w_t$ and sidewall angle $\alpha$ are difficult to change in a controlled manner after the AGPM has been fabricated, the grating depth $h$ can be tuned. Here we demonstrate two techniques for either making the AGPM grating deeper or shallower.

\begin{figure}
\centering
\includegraphics[width=8.8cm]{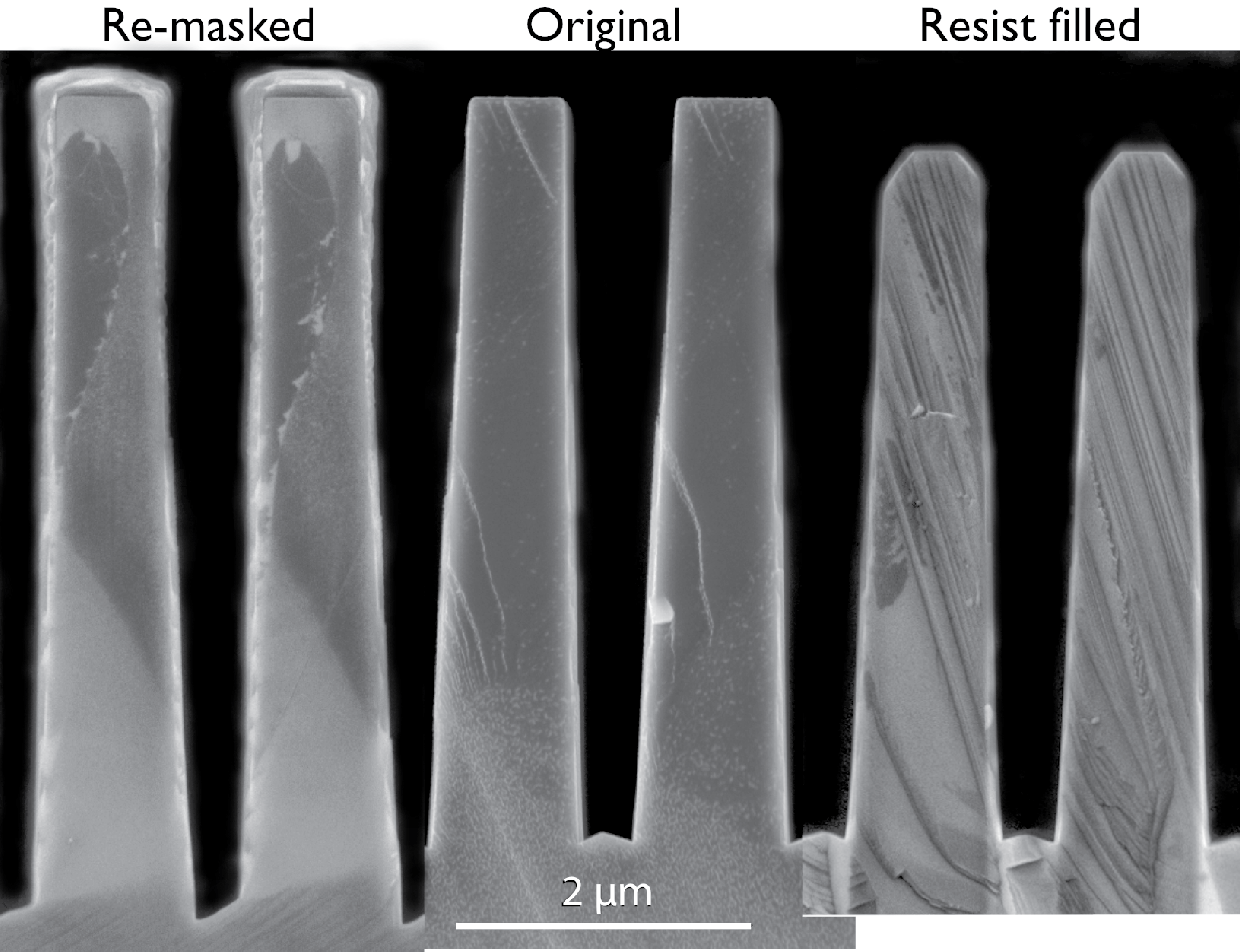}
\caption{Etch profile for the cracked AGPM-L10. \textit{Left.} Profile with deeper grooves, where some remaining Al can be seen on the top of the grating (and upper part of the sidewalls of the grating). \textit{Center.} Grating before tuning. \textit{Right.} Profile with shallower grooves, where the top has become faceted.}
\label{fig:tuning_sem}
\end{figure}

To make the AGPM deeper, we used a technique that we have recently developed for increasing the depth of an already fabricated high aspect ratio diamond grating \citep{Vargas16}. In short, a layer of Al is deposited on top of the diamond AGPM (Fig.~\ref{fig:tuning_process}a), and due to shadowing effects, the top of the grating is covered with a thicker layer than the area at the bottom of the grating. The thin Al layer at the bottom is etched away using ICP-RIE (Fig.~\ref{fig:tuning_process}b), leaving the bottom of the groove without Al and the top still covered with Al. Finally, the AGPM is shortly diamond etched using ICP-RIE (Fig.~\ref{fig:tuning_process}c).

\begin{figure*}
\centering
\includegraphics[width=18cm]{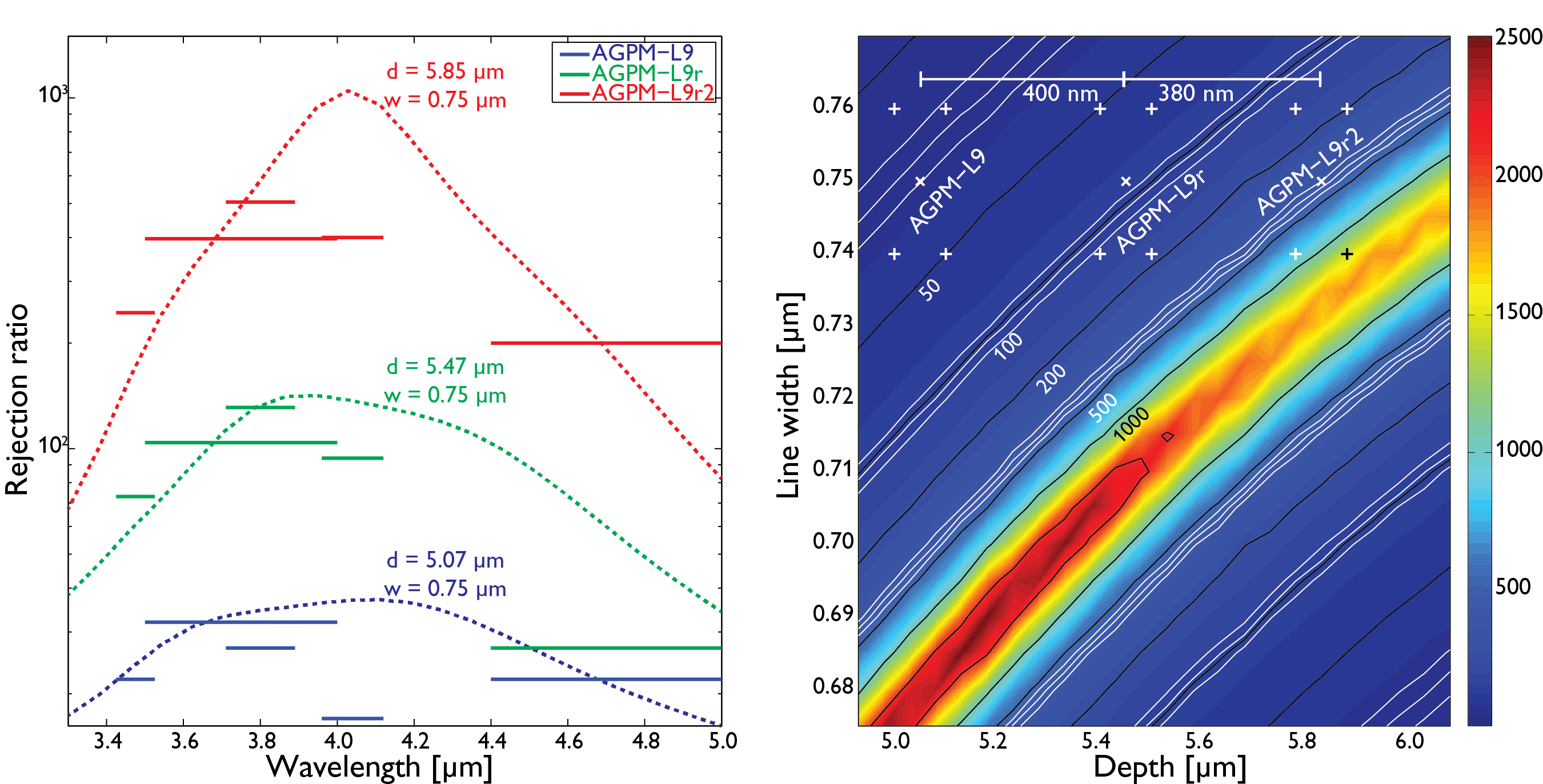}
\caption{Comparison of rejection ratio between L-band and M-band optical measurements and the RCWA model. \textit{Left}. Measured rejection ratio in five different filters for AGPM-L9 and its remastered versions. AGPM-L9r and AGPM-L9r2 were etched 400~nm and 780~nm deeper than the original AGPM-L9, respectively. \textit{Right}. Possible parameter solutions (marked with crosses) for each optimization, overlaid on predicted coronagraphic performance from RCWA simulations (white lines) with $\alpha$=$2\fdg45$ and wavelength region between 3.5-4.0~$\mu$m.}
\label{fig:L5perf}
\end{figure*}

To make the AGPM shallower, a new technique was developed. The grating was filled with photoresist before etching the diamond as above. The diamond surface is hydrophilic due to the previous diamond plasma etching in oxygen chemistry and, in addition, the grating structure makes the surface even more hydrophilic \citep{Karlsson10}; resist dropped on the AGPM will thus immediately fill up the grating. The process is as follows: the AGPM was placed on a spinner, and Shipley S1813 photoresist was dropped on the surface to completely cover the surface. Excess resist was removed by spinning the AGPM substrate at 6000~rpm for 30~s, which leaves about 1~$\mu$m resist on top of the grating (Fig.~\ref{fig:tuning_process}d). To completely bake out the solvent, the AGPM was placed on a hot plate at $115~\degr$C for 20~minutes. The AGPM was then shortly diamond etched using ICP-RIE. This process quickly removes the resist on top of the grating (40-60~s), while the resist inside the grating grooves remains much longer and thus protects the grooves and sidewalls of the diamond structure (Fig.~\ref{fig:tuning_process}e). In other words, the top of the diamond grating will be almost directly attacked by the oxygen plasma, while the resist in the grooves will protect these areas of the diamond grating. Although the etch selectivity between diamond and resist is very low (i.e., resist is etched much faster than diamond), the top diamond area of the grating can be etched several hundreds nanometers before the resist in the grooves was etched away. If there is a need for even shallower grating, the process can be repeated. However, if the grating is etched for too long, faceting of the top of the grating might start to reduce the optical performance (see Fig. 6, right).

AGPM-L10 was used as a test sample to validate our processes to etch deeper and shallower gratings. It was cracked in two halves and characterized (Fig.~\ref{fig:tuning_sem}).
The half that was etched deeper was sputtered with 400~nm thick Al followed by Al plasma etching using Cl$_2$ and BCl$_3$ (gas flows of 15~sccm and 50~sccm, respectively) at 5~mTorr and with an ICP power of 600~W and a bias power of 30~W for 25~s. The diamond substrate was then plasma etched in an oxygen plasma at 5~mTorr with an ICP power of 850~W and a bias power of 220~W for 150~s, resulting in 400~nm deeper grooves. The final grating grooves have a slightly higher sidewall angle, which must be taken into consideration when using RCWA simulations for finding the optimal depth. The other half of AGPM-L10, chosen as a test sample to reduce the grating depth, was filled with photoresist as described above. The process was ended with the same oxygen plasma recipe (and time) as when etching the grating deeper, resulting in 400~nm shallower grooves.

\section{Performance after re-etching} \label{sec:newperf}

\begin{table}
\caption{Rejection ratios in the broadband L filter for the optimized AGPMs.} 
\label{tab:agpm_perf}
\centering          
\begin{tabular}{l c c c c} 
\hline\hline
 & Tuning & $R$ before & $\Delta h$ & $R$ after \\
\multicolumn{1}{c}{Name} & process & tuning & [$\mu$m] & tuning \\
\hline                    
AGPM-L9r & Al deposition & 30 & $+0.40$ & 100 \\ 
AGPM-L9r2 & Al deposition & 100 & $+0.38$ & 400 \\ 
AGPM-L11r & Resist filling & 70 & $-0.32$ & 910 \\ 
AGPM-L12r & Resist filling & 70 & $-0.42$ & 470 \\ 
AGPM-L13r & Resist filling & 110 & $-0.29$ &190 \\ 
\hline                  
\end{tabular}
\end{table}

\begin{figure*}
\centering
\includegraphics[width=11.3cm]{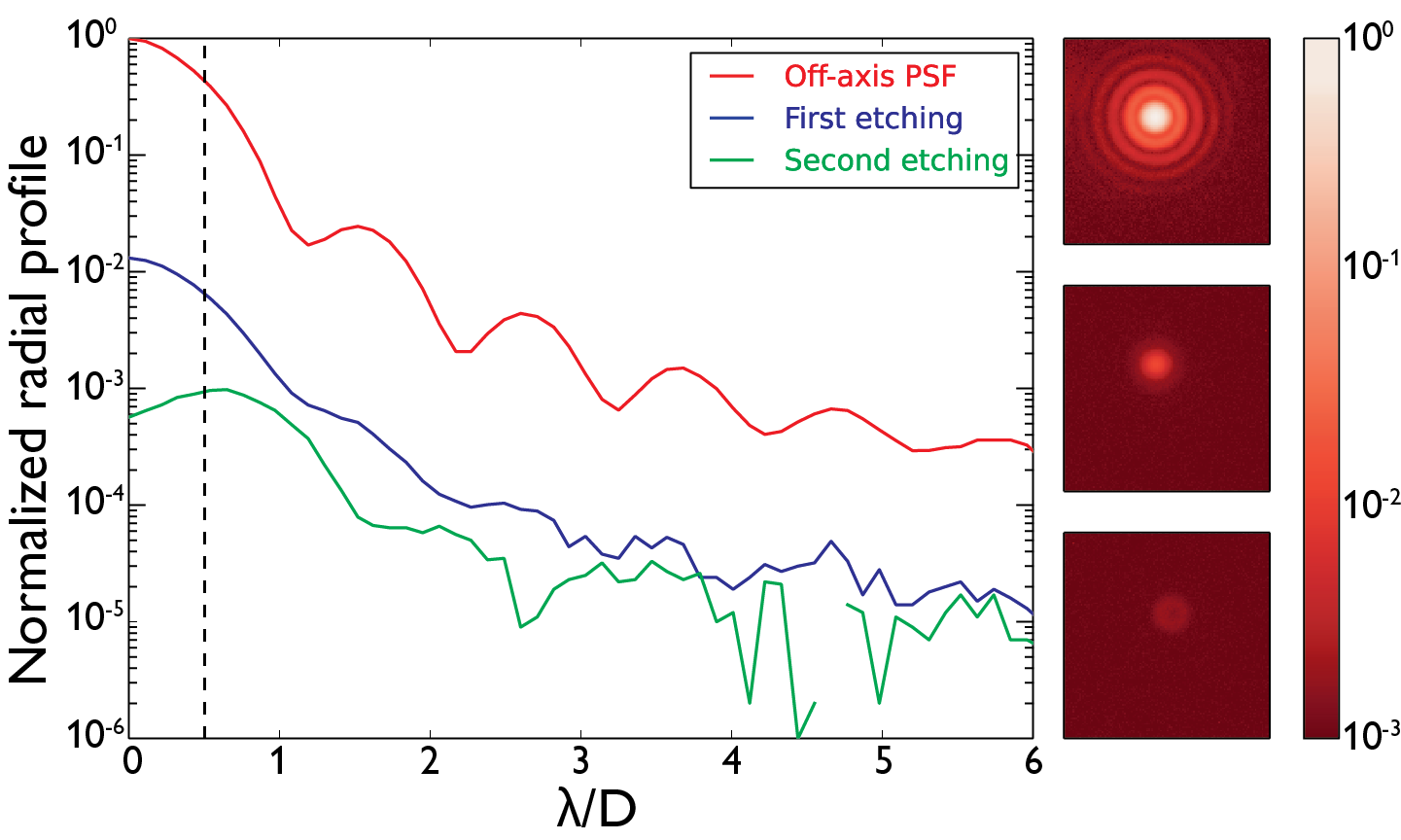} \hspace*{2mm}
\includegraphics[width=6.4cm]{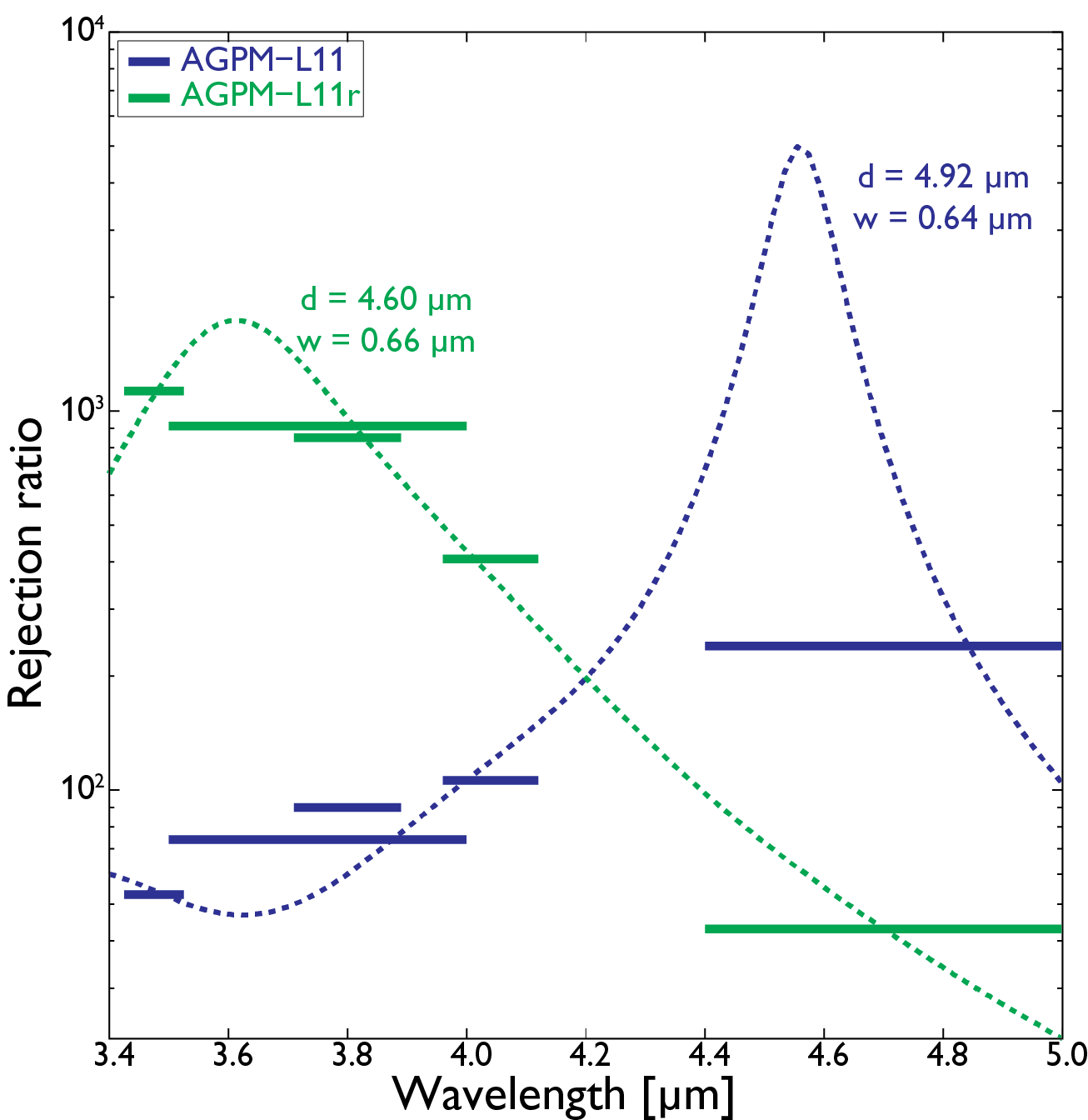}
\caption{Performance of AGPM-L11 before and after the grating tuning process. \textit{Left}. PSF radial profiles measured on the YACADIRE bench in the broadband L filter, together with an illustration of the image shape on the right-hand side. \textit{Right}. Illustration of the experimental coronagraphic performance measured in the five broad- and narrow-band filters, overlaid on the predicted performance based on our best-fit RCWA model of the grating.}
\label{fig:L7perf}
\end{figure*}

The AGPMs showing a rejection ratio around or below 100 (see Table~\ref{tab:agpm_params}) were chosen to test and validate our new tuning processes. Here, we describe the tuning of AGPM-L9 as an example. The experimental rejection ratios measured after initial etching are shown in Fig.~\ref{fig:L5perf} (left) for the broadband L and the three narrow-band filters. These rejection ratios were fit using our RCWA model, giving rise to a family of possible solutions for the line width $w_t$ and grating depth $h$ (thin white lines in Fig.~\ref{fig:L5perf}, right), assuming a sidewall angle $\alpha$ of $2\fdg45$. Thanks to the SEM measurements performed on the twin sample (see Sect.~\ref{sec:fabrication}), we can further constrain the grating parameters, which must be located within the rectangle formed by the four white crosses in Fig.~\ref{fig:L5perf} (right), taking into account the SEM measurement uncertainties of $\pm 10$~nm in $w_t$ and $\pm 50$~nm in $h$. The white lines in the bottom right corner are unrealistic RCWA solutions, as the twin sample can confirm that we have not etched that deep. AGPM-L9 was then processed in the same way as AGPM-L10, with the aim of making the grating 400~nm deeper. The process resulted in a ``new'' phase mask referred to as AGPM-L9r, and the subsequent performance evaluation on our coronagraphic test bench showed an increase of rejection ratio to about 100 in the broadband L filter, as illustrated in Fig.~\ref{fig:L5perf} (left). Based on further RCWA simulations, it was evident that the tuned AGPM was still too shallow, by about 400~nm. A second etch iteration was performed (etch time 200~s) to increase the grating depth by that amount, resulting in the final AGPM-L9r2. The final performance evaluation shows an improvement of the rejection ratio to 400 in the broadband L filter (3.5--4.0~$\mu$m), see Fig.~\ref{fig:L5perf} and Table~\ref{tab:agpm_perf}.

Other successful grating tuning examples include AGPM-L11, AGPM-L12, and AGPM-L13. These AGPMs were originally etched deeper than the previous samples, in an attempt to demonstrate our capability to produce AGPMs delivering good performance simultaneously across the L  and M bands (M-band operations requiring deeper gratings), as shown in Fig.~\ref{fig:rejection_profile}. AGPM-L13 revealed to be the closest approach to a science-grade LM-band AGPM, delivering rejection ratios higher than 100 in both broadband filters. After performance evaluation on our coronagraphic test bench, it was decided to use these three AGPMs to demonstrate the grating depth reduction process. The three AGPMs were thus made shallower using the recipe tested on AGPM-L10. In order to optimize their depth, AGPM-L11 was etched for 120~s, AGPM-L12 for 150~s and AGPM-L13 for 110~s. This corresponded to a decrease of etch depth by 320~nm, 420~nm and 290~nm, respectively. As expected, the three tuned AGPMs all showed better performance in rejection ratio at L band (at the expense of degraded performance at M band), with AGPM-L11 setting the record broadband rejection ratio of 910 (see Table~\ref{tab:agpm_perf}). The coronagraphic PSF of AGPM-L11 in the broadband L filter before and after tuning is shown as an illustration in Fig.~\ref{fig:L7perf}, together with a graphical illustration of its performance in all broad- and narrow-band filters compared to the best-fit RCWA model. A rejection ratio up to 1100 is measured in the short-wave narrow-band filter. At this level of performance, we expect that the optical quality of the wavefront delivered by the YACADIRE test bench becomes a limitation to the measured performance. This is further suggested by the ``donut'' shape of the PSF shown as an inset in Fig.~\ref{fig:L7perf}, which is the expected behavior of the vortex phase mask in presence of low-order aberrations. The actual performance of this AGPM could thus be even better than what is shown here. Based on the measured peak rejection of 900 in broadband L, we would expect a raw contrast of about $10^{-5}$ at an angular separation of $2\lambda/D$ for a perfect wavefront on a circular aperture, rather than the $6\times10^{-5}$ shown in Fig.~\ref{fig:L7perf}. We also note that the noise floor of YACADIRE in broadband L corresponds to a raw contrast of about $2\times10^{-5}$, due to the limited amount of photons making it through the single-mode fiber.

In summary, both methods for optimizing the grating depth (shallower or deeper) were successfully performed. All tuned AGPMs (L9r2, L11r, L12r, and L13r) show better rejection performance. The results for all the tuned AGPMs are summarized in Table~\ref{tab:agpm_perf}, where the suffix ``r'' denotes remastered AGPMs. We note that these sub-micron scale high aspect ratio gratings are never perfect; the angle of the sidewall is not completely constant, and there are so-called trenching effects, which means that the floor of the grating is not at the exact same level everywhere (i.e., not uniform etch depth, see Fig.~\ref{fig:tuning_sem}). A completely accurate RCWA simulation of the AGPM is therefore not possible, but based on the experimental characterization of the AGPM, we can nevertheless optimize the depth of the grating in an efficient way. As long as the grating parameters are reasonably within specification ($\pm10$\%, which is valid for our described manufacturing process), it is always possible to hit an optimal etch depth giving a rejection ratio of 500 or more, which makes it suitable for installation in current and future ground-based infrared high-contrast imagers.

\section{Conclusions and outlook}

Over the last few years, we have produced several AGPMs designed for the thermal infrared regime by etching concentric subwavelength gratings into synthetic diamond substrates. Some of them are now installed in world-leading ground-based observatories, such as the Very Large Telescope, the Large Binocular Telescope and the Keck Observatory. Over the years, we have however discovered that it is very difficult to fabricate AGPMs with good optical performance in a one-iteration process. Errors in the grating parameters will always exist when fabricating nanometer-sized high aspect ratio structures over a relatively large area (cm), thus resulting in degraded optical performance.

In this paper, we have successfully demonstrated that we can tune the AGPM grating depth, that is make it deeper or shallower, even after completing the initial etching process. For that purpose, we have combined the information from SEM micrographs, RCWA modeling, and coronagraphic performance characterization to determine the grating parameters (line width, groove depth, and sidewall angle) with a sufficient accuracy to precisely determine by how much the grating depth needs to be modified to reach the highest possible coronagraphic performance. Two different processes have been presented and validated to reduce or increase the grating depth, enabling the production of L-band AGPMs with broadband rejection ratios up to about 1000:1. Such performance would allow raw contrasts up to $10^{-5}$ to be reached at two resolution elements from the optical axis for a perfect input wave front on a circular, unobstructed aperture. This will ensure that the intrinsic performance of the AGPM does not significantly affect the on-sky coronagraphic performance for current and upcoming infrared high-contrast thermal infrared imagers, such as the Mid-infrared E-ELT Imager and Spectrograph \citep[METIS,][]{Brandl14}, where wave front aberrations and diffraction from the non-circular input pupil will be setting the limit on the achievable raw contrast.

Future work will focus on two main aspects. First, we are in the process of building a new coronagraphic bench \citep[VODCA,][]{Jolivet14}, which should allow the characterization of vortex phase masks in the thermal infrared with a higher dynamic range and better optical quality than currently possible on the YACADIRE bench. Second, we are trying to reduce the grating period down to a sub-micron size to enable operations at K- and H-bands, with promising results already obtained at K-band. For such small grating periods, the errors in the fabrication process will become even more critical compared with L-band AGPMs. Tuning the grating of these AGPMs will certainly be a must to achieve very high rejection performance.

\begin{acknowledgements}
The authors are grateful to J\'er\^ome Parisot (LESIA, Observatoire de Paris), who manages the YACADIRE test bench, for his availability and help during every AGPM test campaigns. The research leading to these results has received funding from the European Research Council under the European Union's Seventh Framework Programme (ERC Grant Agreement n. 337569), the French Community of Belgium through an ARC grant for Concerted Research Action, and the Swedish Research Council (VR) through project grant 621-2014-5959.
\end{acknowledgements}

\bibliographystyle{aa} 
\bibliography{high_perfo_AGPM_rev2} 

\begin{thebibliography}{19}
\expandafter\ifx\csname natexlab\endcsname\relax\def\natexlab#1{#1}\fi

\bibitem[{{Boccaletti} {et~al.}(2008){Boccaletti}, {Carbillet}, {Fusco},
  {Mouillet}, {Langlois}, {Moutou}, \& {Dohlen}}]{Boccaletti08}
{Boccaletti}, A., {Carbillet}, M., {Fusco}, T., {et~al.} 2008, in \procspie,
  Vol. 7015, Adaptive Optics Systems, 70156E

\bibitem[{{Brandl} {et~al.}(2014){Brandl}, {Feldt}, {Glasse}, {Guedel},
  {Heikamp}, {Kenworthy}, {Lenzen}, {Meyer}, {Molster}, {Paalvast}, {Pantin},
  {Quanz}, {Schmalzl}, {Stuik}, {Venema}, \& {Waelkens}}]{Brandl14}
{Brandl}, B.~R., {Feldt}, M., {Glasse}, A., {et~al.} 2014, in \procspie, Vol.
  9147, Ground-based and Airborne Instrumentation for Astronomy V, 914721

\bibitem[{{Defr{\`e}re} {et~al.}(2014){Defr{\`e}re}, {Absil}, {Hinz}, {Kuhn},
  {Mawet}, {Mennesson}, {Skemer}, {Wallace}, {Bailey}, {Downey}, {Delacroix},
  {Durney}, {Forsberg}, {Gomez}, {Habraken}, {Hoffmann}, {Karlsson},
  {Kenworthy}, {Leisenring}, {Montoya}, {Pueyo}, {Skrutskie}, \&
  {Surdej}}]{Defrere14}
{Defr{\`e}re}, D., {Absil}, O., {Hinz}, P., {et~al.} 2014, in \procspie, Vol.
  9148, Adaptive Optics Systems IV, 91483X

\bibitem[{{Delacroix} {et~al.}(2013){Delacroix}, {Absil}, {Forsberg}, {Mawet},
  {Christiaens}, {Karlsson}, {Boccaletti}, {Baudoz}, {Kuittinen}, {Vartiainen},
  {Surdej}, \& {Habraken}}]{Delacroix13}
{Delacroix}, C., {Absil}, O., {Forsberg}, P., {et~al.} 2013, \aap, 553, A98

\bibitem[{{Delacroix} {et~al.}(2012{\natexlab{a}}){Delacroix}, {Absil},
  {Mawet}, {Hanot}, {Karlsson}, {Forsberg}, {Pantin}, {Surdej}, \&
  {Habraken}}]{Del12b}
{Delacroix}, C., {Absil}, O., {Mawet}, D., {et~al.} 2012{\natexlab{a}}, in
  \procspie, Vol. 8446, Ground-based and Airborne Instrumentation for Astronomy
  IV, 84468K

\bibitem[{{Delacroix} {et~al.}(2012{\natexlab{b}}){Delacroix}, {Forsberg},
  {Karlsson}, {Mawet}, {Absil}, {Hanot}, {Surdej}, \& {Habraken}}]{Delacroix12}
{Delacroix}, C., {Forsberg}, P., {Karlsson}, M., {et~al.} 2012{\natexlab{b}},
  \ao, 51, 5897

\bibitem[{{Foo} {et~al.}(2005){Foo}, {Palacios}, \& {Swartzlander}}]{Foo05}
{Foo}, G., {Palacios}, D.~M., \& {Swartzlander}, Jr., G.~A. 2005, Opt. Lett.,
  30, 3308

\bibitem[{Forsberg \& Karlsson(2013)}]{Forsberg13}
Forsberg, P. \& Karlsson, M. 2013, Diamond Relat. Mater., 34, 19

\bibitem[{Gu {et~al.}(2004)Gu, Choi, Liu, Griffin, Girkin, Watson, Dawson,
  McConnell, \& Gurney}]{Gu2004}
Gu, E., Choi, H., Liu, C., {et~al.} 2004, Appl. Phys. Lett., 84, 2754

\bibitem[{Hwang {et~al.}(2004)Hwang, Saito, \& Fujimori}]{Hwang04}
Hwang, D., Saito, T., \& Fujimori, N. 2004, Diamond Relat. Mater., 13, 2207

\bibitem[{{Jolivet} {et~al.}(2014){Jolivet}, {Piron}, {Huby}, {Absil},
  {Delacroix}, {Mawet}, {Surdej}, \& {Habraken}}]{Jolivet14}
{Jolivet}, A., {Piron}, P., {Huby}, E., {et~al.} 2014, in \procspie, Vol. 9151,
  Advances in Optical and Mechanical Technologies for Telescopes and
  Instrumentation, 91515P

\bibitem[{{Karlsson} {et~al.}(2010){Karlsson}, {Forsberg}, \&
  {Nikolajeff}}]{Karlsson10}
{Karlsson}, M., {Forsberg}, P., \& {Nikolajeff}, F. 2010, Langmuir, 26, 889

\bibitem[{{Karlsson} \& {Nikolajeff}(2003)}]{Karlsson03}
{Karlsson}, M. \& {Nikolajeff}, F. 2003, Opt. Express, 11, 502

\bibitem[{Kim {et~al.}(1997)Kim, Xia, Zhao, \& Whitesides}]{Kim97}
Kim, E., Xia, Y., Zhao, X.~M., \& Whitesides, G.~M. 1997, Advanced Materials,
  9, 651

\bibitem[{{Mawet} {et~al.}(2013){Mawet}, {Absil}, {Delacroix}, {Girard},
  {Milli}, {O'Neal}, {Baudoz}, {Boccaletti}, {Bourget}, {Christiaens},
  {Forsberg}, {Gonte}, {Habraken}, {Hanot}, {Karlsson}, {Kasper}, {Lizon},
  {Muzic}, {Olivier}, {Pe{\~n}a}, {Slusarenko}, {Tacconi-Garman}, \&
  {Surdej}}]{Mawet13}
{Mawet}, D., {Absil}, O., {Delacroix}, C., {et~al.} 2013, \aap, 552, L13

\bibitem[{{Mawet} {et~al.}(2005{\natexlab{a}}){Mawet}, {Riaud}, {Absil}, \&
  {Surdej}}]{Mawet05a}
{Mawet}, D., {Riaud}, P., {Absil}, O., \& {Surdej}, J. 2005{\natexlab{a}},
  \apj, 633, 1191

\bibitem[{{Mawet} {et~al.}(2005{\natexlab{b}}){Mawet}, {Riaud}, {Surdej}, \&
  {Baudrand}}]{Mawet05b}
{Mawet}, D., {Riaud}, P., {Surdej}, J., \& {Baudrand}, J. 2005{\natexlab{b}},
  \ao, 44, 7313

\bibitem[{{Mawet} {et~al.}(2010){Mawet}, {Serabyn}, {Liewer}, {Burruss},
  {Hickey}, \& {Shemo}}]{Mawet10}
{Mawet}, D., {Serabyn}, E., {Liewer}, K., {et~al.} 2010, \apj, 709, 53

\bibitem[{{Vargas Catalan} {et~al.}(2016){Vargas Catalan}, {Forsberg}, {Absil},
  \& {Karlsson}}]{Vargas16}
{Vargas Catalan}, E., {Forsberg}, P., {Absil}, O., \& {Karlsson}, M. 2016,
  Diamond Relat. Mater., 63, 60

\end{thebibliography}

\end{document}